\documentclass[11pt,a4paper]{article}
\usepackage{jheppub}
\usepackage{amsfonts} 

\newcommand{\eps}{\epsilon}
\DeclareMathOperator{\Tr}{Tr}

\title{Reduced Massive Gravity with Two St\" uckelberg Fields}
\author{Lasma Alberte$^{a,}$\footnote{Email: \texttt{lasma.alberte@physik.lmu.de}}, Andrei Khmelnitsky$^{a,}$\footnote{Email: \texttt{khmelnitskiy@physik.lmu.de}}}
\affiliation{$ ^a$ Arnold Sommerfeld Center for Theoretical Physics,\\ Ludwig Maximilians University, Theresienstr. 37, 80333 Munich, Germany}

\abstract{We consider the non-linear massive gravity as a theory of a number of St\"uckelberg scalar fields minimally coupled to the Einstein-Hilbert gravity and argue that the counting of degrees of freedom can be done for scalar theory and gravity separately. In this paper we investigate the system with only two St\"uckelberg scalar fields. In this case we find the analytic expression for the determinant of the kinetic matrix of the scalar field Lagrangian and perform the full constraint analysis. In $1+1$ space-time dimensions the theory corresponds to the full non-linear massive gravity, and this determinant vanishes identically. In this case we find two first-class constraints, and present the corresponding gauge symmetry of the theory which eliminates both scalar degrees of freedom. In $3+1$ dimensions  the determinant of the kinetic matrix does not vanish identically and, for generic initial conditions, both scalar fields are propagating. 
}
\subheader{LMU-ASC 15/13}

\begin{document}
\maketitle
\flushbottom

\section{Introduction}
The observation of the accelerated expansion of our universe is the driving motivation for various infrared modifications of general relativity. One of the theoretically most natural infrared modification would be to give a small mass to the graviton. Since the early discovery of the quadratic Fierz-Pauli mass term for metric perturbations in \cite{pauli}, there has been an ongoing search for a healthy non-linear completion of massive gravity. The construction of the non-linear graviton mass term is based on the use of an auxiliary non-dynamical reference metric, which as an absolute object would break the diffeomorphism invariance of general relativity. The diffeomorphism invariance can be restored by introducing four St\"uckelberg scalars, corresponding to the four coordinate transformations \cite{Siegel:1993sk,arkani,mukh}. However, a generic theory of four St\"uckelberg scalars together with the two degrees of freedom of massless graviton propagates six degrees of freedom in total. It is one degree of freedom too much in comparison to the five degrees of freedom expected from the massive spin-2 representations of the Poincar\'e group. Moreover, the additional degree of freedom is sick and represents the (in)famous Boulware-Deser (BD) ghost~\cite{boul}.

After an order-by-order construction of a non-linear theory which is ghost-free in the decoupling limit in \cite{gr}, a full resummed theory of non-linear massive gravity was proposed by de Rham, Gabadadze, and Tolley (dRGT) \cite{grt}. In unitary gauge this theory has been shown to propagate five degrees of freedom~\cite{Hassan:2011hr,Hassan:2011ea}. The Hamiltonian analysis of the full diffeomorphism invariant theory including the four St\"uckelberg fields also seems to confirm the expectation that the dRGT theory propagates at most five degrees of freedom \cite{deRham:2011rn,Hassan:2012qv,Kluson:2012wf} (for recent counterarguments see \cite{Chamseddine:2013lid}). However, the canonical analysis of dRGT theory in the presence of the four scalar fields is intricate, and in the existent literature it is often obscured either by mixing the gravitational and scalar degrees of freedom or by introduction of new auxiliary fields.

In the present paper we take a different point of view and treat dRGT massive gravity as a theory of St\"uckelberg scalar fields $\phi^A$ coupled to the Einstein-Hilbert gravity. Since the theory is reparametrization invariant, and the scalars are coupled to gravity minimally, we shall count the degrees of freedom propagated by the metric and by the scalar fields separately. Hence the absence of the sixth mode in dRGT theory should manifest itself as the feature of the scalar fields Lagrangian alone.

Motivated by these considerations we study the dynamics of the St\"uckelberg scalar fields given by the dRGT mass term \cite{grt}. We observe that, if seen as a particular scalar field theory, the dRGT scalar field Lagrangian allows for an arbitrary number of scalar fields in it. In particular, the number of scalar fields $N$ can be chosen to be less than the space-time dimension $d+1$ without affecting the diffeomorphism nor the space-time Lorentz invariance of the theory. We dub the dRGT theories of gravity with reduced number $N<d+1$ of St\" uckelberg scalar fields as ``reduced massive gravity".

The simplest particular cases of such dRGT inspired scalar theories include, for $d = 0$, the action of a massive relativistic particle in $N$ dimensions and, for $N =1$, the single ``k-essence'' field with DBI-like action \cite{ArmendarizPicon:1999rj}. Another ``simple" choice is arbitrary $N$ fields in $1+1$ dimensions, and gives the action of a relativistic string in $N$-dimensional target space-time. In the case $N=3$, with three scalar fields living in a configuration space diffeomorphic to $\mathbb R^3$, the reduced dRGT action can be regarded as a particular effective field theory of homogeneous solid \cite{Dubovsky:2005xd}. The degree of symmetry of the solid depends on the isometries of the metric $f_{AB}(\phi)$ in the internal space of scalar fields. If the metric is symmetric under the $SO(3)$ group, and the action contains only the term, invariant under the volume preserving diffeomorphisms, then it describes a perfect fluid. The case with the number of scalar fields $N\geq d+1$ has been recently discussed in \cite{Gabadadze:2012tr,Andrews:2013ora} as a theory of multiple Galileon fields covariantly coupled to the dRGT massive gravity.

In general, the solutions of the reduced massive gravity theories are expected to break Lorentz and rotational symmetries and lead to anisotropic cosmologies. The pattern of such breaking is determined by the number of scalar fields and the signature and isometries of the reference metric. The connection of reduced massive gravity theories to the Lorentz violating massive gravity theories will be discussed in more detail in the main body of the paper. Another possible application of reduced massive gravity theories could be found in modeling the translational symmetry breaking and momentum dissipation in holography. In particular, in~\cite{Vegh:2013sk} the conductivity in the boundary theory was calculated in the presence of a Lorentz violating graviton mass term in the bulk, that originated from the dRGT-like action with two St\"uckelberg fields and Euclidean reference metric. The models discussed in our paper could be further used in holographic constructions.

In this paper we consider the case of reduced massive gravity with two St\" uckelberg fields. It is the simplest case with several scalar fields involved, in which we can write the Hamiltonian and constraint structure explicitly. We perform the full Hamiltonian analysis of the scalar field sector and find that, in distinction from the dRGT massive gravity the determinant of the kinetic matrix does not vanish. Hence the scalar field Lagrangian in general propagates two degrees of freedom. We formulate the condition for the scalar field configurations on which the determinant vanishes and investigate the different regions in the phase space of scalar fields. We show that on the singular surface, where the determinant of the kinetic matrix vanishes, the theory is equivalent to $1+1$-dimensional massive gravity and thus has no dynamical degrees of freedom. We also show that the regular solutions away but in close vicinity of the singular surface approach the singular surface but can never reach it in finite time. At the same time any perturbation of the singular solution drives the system away from this singular surface. In quantum theory the vanishing of the determinant signals the strong coupling regime for the scalar fields, and the dynamics in the vicinity of the singular surface are highly affected by quantum corrections. Whether or not the two dynamical degrees of freedom away from the singular surface contain ghost modes might depend on the particular choice of the reference metric in the configuration space of the scalar fields. We do not address this question in the present paper, but leave it for future studies.

The paper is organized as follows. In section \ref{sec:2} we recall the formulation of dRGT massive gravity. In section \ref{sec:3} we formulate the theory of reduced massive gravity and perform the Hamiltonian analysis away from the singularity surface. In section~\ref{sec:4} we consider the behaviour of the system on the singular surface, and show that it is equivalent to $1+1$ dimensional massive gravity. We perform the canonical analysis in this case and find the gauge symmetry of the scalar fields, eliminating both scalar degrees of freedom. Section~\ref{sec:5} is devoted to conclusions.

\section{Non-linear massive gravity in St\"uckelberg formulation}\label{sec:2}

The non-linear massive gravity action can be written in terms of the variables
\begin{equation}
\mathcal K^\mu_\nu=\delta^\mu_\nu-\left(\sqrt{g^{-1}f}\right)^\mu_\nu\;,
\end{equation}
where $g^{\mu\nu}$ is the inverse space-time metric, and $f_{\mu\nu}$ is an auxiliary reference metric. The full dRGT action is given by
\begin{equation}\label{act0}
\mathcal L_{EH}+m^2\mathcal L_\phi=\frac{M^2_P}{2}\sqrt{-g}R+m^2\sqrt{-g}\sum_{n=0}^4\tilde \alpha_n\mathsf e_n(\mathcal K)\;,
\end{equation}
where the characteristic polynomials $\mathsf e_n(\mathbb X)$ of a $4\times 4$ matrix $\mathbb X$ are
\begin{align*}
\mathsf e_0(\mathbb X)&=1\;,\qquad\mathsf e_1(\mathbb X)=[\mathbb X]\;,\qquad
\mathsf e_2(\mathbb X)=\frac{1}{2}\left([\mathbb X]^2-[\mathbb X^2]\right)\;,\\
\mathsf e_3(\mathbb X)&=\frac{1}{6}\left([\mathbb X]^3-3[\mathbb X][\mathbb X^2]+2[\mathbb X^3]\right)\;,\qquad \mathsf e_4(\mathbb X)=\det\mathbb X\;.
\end{align*}
The squared brackets denote the traces, and the coefficients $\tilde \alpha_n$ are arbitrary. It is also possible to rewrite the mass term in terms of the characteristic polynomials of the square root matrix $\left(\sqrt{\Omega}\right)^\mu_\nu\equiv \left(\sqrt{g^{-1}f}\right)^\mu_\nu$ as 
\begin{equation}\label{mt}
\mathcal L_\phi=\sqrt{-g}\sum_{k=0}^4\tilde \beta_k\mathsf e_k(\sqrt{\Omega})\;,
\end{equation}
with the coefficients $\tilde \beta_k$ given by
\begin{equation}\label{beta}
\tilde \beta_k=\sum_{n=k}^4(-1)^k \left(\begin{array}{c}
4-k\\
n-k\\
\end{array}\right)\tilde\alpha_n\;.
\end{equation}
The characteristic polynomials of an $n\times n$ matrix $\mathbb X$ can be rewritten as the characteristic polynomials of its eigenvalues $\lambda_i$ as $\mathsf e_n(\mathbb X)=\mathsf e_n(\lambda_i)$ \cite{nieuw,rosen}, where 
\begin{align*}
\mathsf e_0(\lambda_i)&=1\;,\qquad\mathsf e_1(\lambda_i)=\sum_{i}\lambda_i=[\mathbb X]\;,\\
\mathsf e_2(\lambda_i)&=\sum_{i<j}\lambda_i\lambda_j\;,\\
\vdots\\
\mathsf e_n(\lambda_i)&=\lambda_1\lambda_2\dots\lambda_n=\det\mathbb X\;.
\end{align*}
Since here the matrix $\mathbb X=\sqrt{\Omega}$ is a square root matrix, we note that its eigenvalues are, by definition, equal to the square root $\sqrt{\lambda_i}$ of the eigenvalues of the matrix $\Omega$. Hence the mass term \eqref{mt} can be rewritten in terms of the eigenvalues of the matrix $\Omega^\mu_\nu$, without the need of finding the explicit expression of the square root matrix itself, as
\begin{equation}\label{mt2}
\mathcal L_\phi=\sqrt{-g}\sum_{k=0}^4\tilde \beta_k\mathsf e_k(\sqrt{\lambda_i})\;.
\end{equation}
Since the mass term \eqref{mt} explicitly depends on the auxiliary metric $f_{\mu\nu}$ it breaks the diffeomorphism invariance of general relativity. It can be fully restored by introducing four St\" uckelberg scalar fields $\phi^A,\,A=0,1,2,3$, corresponding to the four coordinate transformations as $f_{\mu\nu}=\partial_\mu\phi^A\partial_\nu\phi^B\eta_{AB}$ \cite{arkani}. In addition, in this parametrization the auxiliary metric is invariant under the Lorentz transformations $\Lambda^A_B$ in the scalar field space \cite{mukh}. Hence the scalar field indices $A,\,B$ are raised and lowered with the Minkowski metric $\eta_{AB}=\textrm{diag}\,({}-{}+{}+{}+{})$. In this case the reference metric $f_{\mu\nu}$ is said to be `flat' since it is simply a coordinate transformation from the flat Minkowski metric $\eta_{AB}$. An arbitrary `curved' reference metric $f_{\mu\nu}$ can be obtained by replacing the flat metric $\eta_{AB}$ with some arbitrary scalar field metric $f_{AB}(\phi)$ \cite{Alberte:2011ah}. Then the Lorentz transformations in the scalar field configuration space are replaced by the isometries of the metric $f_{AB}(\phi)$.

The St\" uckelberg formulation of the massive gravity allows for an equivalent form of the mass term \eqref{mt} by introducing a diffeomorphism invariant matrix
\begin{equation}
\mathcal I^{A}_B\equiv g^{\mu\nu}\partial_\mu\phi^A\partial_\nu\phi^Cf_{BC}\;.
\end{equation}
Since the traces and eigenvalues of the matrices $\mathcal I^A_B$ and $\Omega^\mu_\nu=g^{\mu\rho}\partial_\rho\phi^A\partial_\nu\phi^Bf_{AB}$ are equal then the mass term \eqref{mt2} can be equivalently written in terms of the eigenvalues of $\mathcal I^A_B$. This rewriting makes manifest that any non-linear massive gravity theory can be viewed as a theory of a number of scalar fields minimally coupled to gravity.

\section{Reduced massive gravity}\label{sec:3}

In the present paper we adopt the point of view that the mass term Lagrangian $\mathcal L_\phi$ used in the non-linear dRGT massive gravity is a Lagrangian describing four St\"uckelberg scalar fields coupled to gravity. The motivation of restricting the number of scalar fields to the number of space-time dimension in the context of non-linear massive gravity is that around the background solution $g_{\mu\nu}=\eta_{\mu\nu},\,\phi^A=x^\mu\delta^A_\mu$ the metric perturbations have a Lorentz invariant mass term of the Fierz-Pauli form at the quadratic level. However, if seen as describing a theory of scalar fields, the action 
\begin{align}\label{act}
\mathcal L_\phi=\sqrt{-g}\sum_{n=0}^4\alpha_n\mathsf e_n(\mathbb I -\sqrt{\mathcal I}),\qquad
\mathcal I^{A}_B\equiv g^{\mu\nu}\partial_\mu\phi^A\partial_\nu\phi^Cf_{BC}(\phi)
\end{align}
describes just some particular theory of derivatively coupled scalar fields, and depends only on their first derivatives. This theory is diffeomorphism invariant even when the number of scalar fields is not equal to the space-time dimension. Therefore from the scalar field theory point of view the number of scalar fields $N$ can be chosen arbitrary, both less or greater than $d+1$. In the case $N \neq d+1$ the matrices $\mathcal I^A_B$ and $\Omega^\mu_\nu$ have different dimensions, $N\times N$ and $(d+1)\times (d+1)$ respectively. Nevertheless, the non-vanishing eigenvalues of these matrices are equal, and both formulations~\eqref{act0} and~\eqref{act} of the action are still equivalent, even though the formulation in terms of the smaller matrix is evidently simpler.\footnote{The coefficients $\tilde\alpha_n,\,\alpha_n$ in~\eqref{act0} and~\eqref{act} respectively coincide only when the number of fields equals the space-time dimension, i.e. when $N=d+1$.} 

In this work we focus on the case of two scalar fields $\phi^A=\{\phi^0,\,\phi^1\}$ in $3+1$ dimensions as the simplest non-trivial case inspired by the dRGT massive gravity. Using the diffeomorphism invariant variables $\mathcal I^A_B$ proves to be particularly useful for this setup since the matrix $\mathcal I^A_B$ is a $2\times 2$ matrix in this case whereas $\Omega^\mu_\nu$ in a $(3+1)$-dimensional space-time is a $4\times 4$ matrix. The action of the scalar fields in any $d\geq 1$ then takes the simple form 
\begin{equation}\label{red}
\mathcal L_\phi=\sqrt{-g}\left(\alpha_0+\alpha_1\mathrm {Tr}(\mathbb I -\sqrt{\mathcal I})+\alpha_2\det(\mathbb I -\sqrt{\mathcal I})\right),
\end{equation}
where we have used the fact that for any $2\times 2$ matrix $\mathbb X$ the polynomials $\mathsf e_{3,4}(\mathbb X)$ vanish. In the case $\alpha_0=\alpha_1=0$ and for the scalar field metric taken to be the Minkowski metric $\eta_{AB}$, the full theory $\mathcal L=\mathcal L_{EH}+m^2\mathcal L_\phi$ has the solution
\begin{equation}\label{sol}
g^{\mu\nu}=\eta^{\mu\nu},\quad \phi^0=x^0,\quad\phi^1=x^1.
\end{equation}
The quadratic action for the perturbations  
\begin{equation}
h^{\mu\nu}\equiv g^{\mu\nu}-\eta^{\mu\nu},\quad \chi^A\equiv \left(\phi^0-x^0\right)\delta^A_0+\left(\phi^1-x^1\right)\delta^A_1
\end{equation}
then reads
\begin{equation}\label{act2}
\mathcal L_\phi^{(2)}=2\left[(h^{01})^2-h^{00}h^{11}\right]+4h^{AB}\left[\eta_{AB}\partial_C\chi^C-\partial_A\chi_B\right]+2\left[(\partial_A\chi^A)^2-\partial^A\chi^B\partial_B\chi_A\right],
\end{equation}
and the indices $A,B = 0,1$. In $1+1$ space-time dimensions this action coincides with the action for metric and scalar field perturbations around the Minkowski background in massive gravity. However, in $3+1$ space-time dimensions, this corresponds to a Lorentz-violating Fierz-Pauli-type mass term for metric perturbations. A thorough analysis of Lorentz-violating graviton mass terms, preserving the Euclidean symmetry of the three-dimensional space was carried out in \cite{Dubovsky:2004sg} (see also an earlier work \cite{Rubakov:2004eb}). In our case, the symmetry of the three-dimensional rotations is in general not preserved by the ground state of the theory. Therefore, the possible mass terms of our theory go beyond the considerations of \cite{Dubovsky:2004sg}. Out of their investigated mass terms, only the mass term with $m_2=m_3\neq 0,\,m_{0,1,4}=0$ can be obtained in reduced massive gravity (if the number of scalars $N=3$). However, the stability analysis of \cite{Dubovsky:2004sg} does not directly apply to our case since the number of Goldstone fields is different.

A particular instance when the (3+1)-dimensional dRGT theory of massive gravity reduces to the special case of two St\" uckelberg fields is the case of a degenerate reference metric. To see this one can consider the spherically symmetric ansatz $\phi^0=f(t,r)$, $\phi^i=g(t,r),\,i=1,2,3$.\footnote{We note that this is not the ansatz usually studied in the context of the spherically symmetric solutions of dRGT theory. Instead the common ansatz is $\phi^0=g(t,r),\,\phi^i=f(r,t)x^i/r$.} For the flat auxiliary metric 
$f_{\mu\nu}=\partial_\mu\phi^A\partial_\nu\phi^B\eta_{AB}$ in spherical coordinates, this gives a matrix with the only non-zero entries in the upper-left $2\times 2$ matrix, and it can be easily reparametrized by using only two St\" uckelberg fields. This illustrates our point that the reduced massive gravity with the number of scalar fields $N$ less than the space-time dimension is equivalent to dRGT theory with a degenerate reference metric $f_{\mu\nu}$ (or $f_{AB}$ equivalently). However, the spherically symmetric ansatz given above reduces to the degenerate reference metric only in the absence of perturbations.

\subsection{Number of degrees of freedom}
We would now like to estimate the total number of degrees of freedom propagated by the full non-linear theory of gravity and two scalar fields. For this we will use the Dirac's approach to the Hamiltonian analysis of constrained systems \cite{dirac,Henneaux:1990au}.

As was already mentioned, due to the fact that the action \eqref{red} is reparametrization invariant and that the scalar fields are coupled to gravity minimally, i.e. only through the terms $g^{\mu\nu}\partial_\mu\phi^A\partial_\nu\phi^B$, it is legitimate to count the number of degrees of freedom propagated by the scalar field action and the Einstein-Hilbert action separately. In such a diffeomorphism invariant theory of gravity and minimally coupled scalar fields, the Hamiltonian vanishes on the constraint surface, and both the lapse and the shift enter the Hamiltonian linearly. This implies the appearance of in total $2(d+1)$ first-class constraints, which can be used to reduce the number of gravitational degrees of freedom to $d(d+1)/2-2(d+1)$. The dynamics of the scalar fields then shall be generated by the usual Hamiltonian of the scalar field action alone, contained in the Hamiltonian constraint of the full theory. Therefore, the scalar field dynamics in such a theory can be considered separately from gravity. Then naively one would expect that the number of degrees of freedom propagated by any dRGT-type massive gravity in $(d+1)$-dimensional space-time (with $d\geq 2$) equals to
\begin{equation}
\# \textrm { d.o.f. }=\frac{1}{2}(d-2)(d+1)+N
\end{equation}
where the first term accounts for the degrees of freedom propagated by the massless graviton, and the second term is just the number of scalar fields. This naive counting demonstrates why, in (3+1)-dimensional space-time, a general non-linear massive gravity theory with four St\"uckelberg fields propagates six degrees of freedom. It has been demonstrated that in the dRGT subclass of massive gravity theories at most five degrees of freedom propagate due to the special structure of the graviton mass term \eqref{act0} \cite{grt,Hassan:2011hr}. In St\"uckelberg language it is clear that, in order for the assertion to be true, the scalar field Lagrangian \eqref{mt} has to have a very special structure such that it propagates less degrees of freedom than the number of fields. In other words, in the non-linear dRGT massive gravity the four St\"uckelberg fields do not correspond to four independent degrees of freedom \cite{deRham:2011rn} (see also \cite{Hassan:2012qv,Kluson:2012wf}). This can be seen from the vanishing of the determinant of the kinetic (Hessian) matrix of the scalar field Lagrangian  
\begin{equation}\label{kin}
\mathcal A_{AB}\equiv\frac{\partial^2\mathcal L_\phi}{\partial \dot\phi^A\partial\dot\phi^B}\;.
\end{equation}
Hence the equations of motion of the scalar fields are not independent from each other, and there exists (at least) one combination of the equations of motion which gives a constraint equation relating the canonical momenta of the scalar fields. As a result, in dRGT massive gravity the scalar fields propagate at most $N-1=3$ degrees of freedom.

Our ultimate goal is to find the constraint structure of the scalar field part of the full dRGT massive gravity while keeping the space-time metric arbitrary. In this paper we start with the case of the reduced massive gravity \eqref{red} with two scalar fields. For this we explicitly calculate the determinant of the kinetic matrix of the theory. Curiously, we show that the naive expectation, that also in the case of two scalar fields the determinant vanishes and the theory propagates $N-1=1$ degree of freedom, is not met. Instead we find that, in general, the determinant is not equal to zero, and thus there are two dynamical degrees of freedom in the scalar field sector.

\subsection{Determinant of the kinetic matrix}
In the case of two scalar fields, the only non-vanishing characteristic polynomials of the square root matrix can be explicitly expressed in terms of the $\mathrm{Tr}\,\mathcal I$ and $\det\mathcal I$ as
\begin{align}
\mathsf e_0(\sqrt{\mathcal I})=1\;,\qquad \mathsf e_1(\sqrt{\mathcal I})&=\mathrm{Tr}\sqrt{\mathcal I}= \left(\mathrm{Tr}\,\mathcal I+2\sqrt{\det \mathcal I}\right)^{1/2}\;,\\
\mathsf e_2(\sqrt{\mathcal I})&=\det\sqrt{\mathcal I}=\sqrt{\det\mathcal I}\;. 
\end{align}
Then the scalar field action \eqref{red} reads
\begin{equation}\label{act3}
\mathcal L_\phi=\sqrt{-g}\left[\beta_0+\beta_1\left(\mathrm{Tr}\,\mathcal I+2\sqrt{\det \mathcal I}\right)^{1/2}+\beta_2\sqrt{\det\mathcal I}\right].
\end{equation}
Since the $\beta_0$ term does not affect the dynamics of the scalar fields, in what follows we set $\beta_0=0$. We also note that in dRGT massive gravity case, where the number of scalar fields $N$ coincides with the number of space-time dimensions, the highest order term with $\beta_N$ is usually dropped since it is a total derivative. In the reduced massive gravity, however, the term with $\beta_{N=2}$ does contribute to the dynamics of the scalars and, in general, cannot be neglected.

In order to separate the time derivatives of the scalar fields while keeping the space-time metric arbitrary, we employ the ADM formalism \cite{Arnowitt:1962hi}. In ADM variables for the metric components 
\begin{equation}
g^{\mu\nu} =
\left(
\begin{array}{r c}
-\frac{1}{N^2}  & \frac{N^i}{N^2}\\[.6em]
\frac{N^i}{N^2} & \gamma^{ij}-\frac{N^iN^j}{N^2}
\end{array}
\right)
\end{equation}
the matrix $\mathcal I^A_B$ can be expressed as
\begin{equation}
\mathcal I^A_B=\left(-D\phi^AD\phi^C+S^{AC}\right)f_{BC} \;,
\end{equation}
where $D\equiv\frac{1}{N}\left(\partial_0-N^i\partial_i\right)$, and the matrix $S^{AC}\equiv \gamma^{ij}\partial_i\phi^A\partial_j\phi^C$ depends only on the spatial derivatives of the scalar fields.
The canonical momenta conjugated to the scalar fields are given by
\begin{align}
\pi_A\equiv\frac{1}{N}\frac{\partial\mathcal L_\phi}{\partial D\phi^A} &= -\sqrt\gamma \left(\frac{\beta_1}{\left(\Tr\mathcal I+2\sqrt{\det\mathcal I}\right)^{1/2}}\left[D\phi_A+\frac{1}{\sqrt{\det\mathcal I}}\left(  S\,f_{AB}- S_{AB} \right) D\phi^B\right] + \right. \nonumber\\
&\qquad \left.\vphantom{\frac{\beta_1}{\left(\Tr\mathcal I+2\sqrt{\det\mathcal I}\right)^{1/2}}} {}+\frac{\beta_2}{\sqrt{\det\mathcal I}}\left(S\,f_{AB}- S_{AB} \right) D\phi^B \right) \;. \label{eq:pi}
\end{align}
Here $S \equiv \Tr S^A_B$, and the $\Tr \mathcal{I}$ and $\det \mathcal I$ also depend on the time derivatives $D\phi^A$ as
\begin{align*}
\Tr \mathcal{I} &\equiv \Tr \mathcal{I}^A_B = S - D\phi^A D\phi_A\;,\\
\det \mathcal I &\equiv \det \mathcal I^A_B = \det S -  D\phi^A D\phi^B \left( S\,f_{AB}- S_{AB} \right)\;,
\end{align*}
where $\det S\equiv\det S^A_B=\det f\det S^{AB}$. The determinant of the kinetic matrix is given by
\begin{align}
\det \mathcal A^A_{B}&=-\det g\frac{\det S}{(\det\mathcal I)^2}\left(\frac{\beta_1^2}{\left(\mathrm{Tr}\,\mathcal I+2\sqrt{\det\mathcal I}\right)^2}\left[\sqrt{\det\mathcal I}\,\mathcal I^A_B(3S\delta_{A}^{B}-2S_{A}^{B})\right.- \right.\nonumber\\
&\qquad  {} -\mathcal I^{A}_{B}\left(\mathrm{Tr}\,\mathcal I+2S\right)(S_{A}^{B}-S\delta_{A}^{B})-2S\det S\Big]+\nonumber\\
&\qquad  +\frac{\beta_1\beta_2}{\left(\mathrm{Tr}\,\mathcal I+2\sqrt{\det\mathcal I}\right)^{3/2}}\left[\sqrt{\det\mathcal I}\left(S\,\mathrm {Tr}\mathcal I+4\det S\right)-\right.\nonumber\\\label{det}
&\qquad {}- \left(2S\mathcal I^{A}_{B}+S^{A}_{B}\mathrm{Tr}\,\mathcal I\right)\left(S_{A}^{B}-S\delta_{A}^{B}\right)-2S\det S\Big]+\beta_2^2\det S\Bigg)\;.
\end{align}
This expression is valid for any choice of the scalar field metric $f_{AB}(\phi)$ as long as it does not involve the time derivatives of the scalar fields. The determinant depends on temporal (contained within the matrix $\mathcal I^A_B$) and spatial derivatives of the scalar fields. In general it has a non-zero value which depends on the choice of initial conditions. The only special case when the determinant vanishes identically is if we are considering a two-dimensional space-time where the matrix $S^{AB}=\gamma^{11}\partial_1\phi^A\partial_1\phi^B$ is a matrix of rank one, and $\det S\equiv 0$. This case corresponds to the two-dimensional massive gravity and our findings are in agreement with the previous work by de Rham et al. \cite{deRham:2011rn}. If the $\det S$ factor appears also in the theory with four scalar fields, then the full dRGT massive gravity in $3+1$ dimensions also has the identically vanishing kinetic matrix, and thus at most five degrees of freedom in total.

We thus conclude that in general the action \eqref{act3} describes two independent dynamic fields. However, on the surface $\det S=0$ in the configuration space, the Lagrangian equations of motion are degenerate and determine the second time derivative only for one independent combination of fields. In the theory of partial differential equations the solutions that entirely belong to the $\det S=0$ subspace are called \emph{singular solutions}~(cf.~\cite{PDE}). In other words, singular solutions of a system of differential equations are the solutions which belong to the surface where the number of independent highest time derivatives is less than the number of the fields. Such solutions in general are the envelopes of families of regular solutions of the system, and at each fixed moment of time coincide with some regular solution (or the whole family of regular solutions). It means that the initial conditions on this surface do not specify a unique solution since there are other solutions of the theory which are touching the $\det S=0$ surface at the initial moment of time. We note that this discussion holds only classically. In the full quantum theory the vanishing of the determinant of the kinetic matrix signals that the scalar fields are infinitely strongly coupled, and the quantum effects are crucial for the dynamics of the system near the singular surface.

It is interesting to note that the trivial solution \eqref{sol} with $\phi^A=x^A$ is on the surface $\det S=0$. However, any perturbations around this solution defined as $\chi^A=\phi^A-x^A$ will no longer be on the singular surface and will propagate two degrees of freedom. In order to understand the dynamics of such field perturbations it is instructive to study the behaviour of the system in a close vicinity of the singular surface. Note that for $\beta_2\neq 0$ the condition $\det S=0$ is also a necessary condition for $\det \mathcal A=0$. Therefore in our discussion of singular solution we will focus on the singular surface $\det S=0$. Although in general the determinant of the kinetic matrix could vanish also in some other regions of the phase space.

\subsection{Hamiltonian analysis away from $\det S=0$}
For any initial conditions away from the surface $\det S = 0$ the expression~\eqref{eq:pi} for momenta is invertible, and the system contains two propagating degrees of freedom. In order to qualitatively understand the dynamics of the system in the vicinity of the singular surface we  fix the scalar fields metric to be flat  $f_{AB} = \eta_{AB}$ and construct the Hamiltonian for the limiting cases, when only one of the terms in Lagrangian~\eqref{act3} is present. 

First, we consider the case when $\beta_2 = 0$ and $\beta_1 = 1$. The action with only $\beta_1$ term present, in the case of four scalar fields, was already studied as a special case of the dRGT theory and is named as ``minimal non-linear massive gravity". Our Hamiltonian is in agreement with the previous results~(cf.~\cite{Kluson:2012wf}). The expression~\eqref{eq:pi} for the momenta can be inverted to give
\begin{equation}
D{\phi}^A = - \frac{\left[\Tr\mathcal I+2\sqrt{\det\mathcal I}\right]^{1/2}}{\sqrt\gamma\left(1 + \frac{S}{\sqrt{\det\mathcal I}} + \frac{\det S}{\det\mathcal I}\right)} \left( \pi^A + \frac1{\sqrt{\det\mathcal I}}\, S^{AB} \pi_B \right)\,.
\end{equation}
It still does not allow to express the velocities in the terms of momenta completely, but it turns out to be enough in order to obtain the Hamiltonian in terms of $S^{AB}$ and $\pi_A$. After some algebra and with the help of the relation $\frac{\det S}{\det\mathcal I} = 1 + \gamma^{-1} \pi^A\pi_A$, the Hamiltonian takes the following form:
\begin{equation}\label{eq:ham}
\mathcal H = - N\sqrt\gamma\left( S + \gamma^{-1}\,\pi_A \pi_B S^{AB} + 2\sqrt{\det S\, \left(1 + \gamma^{-1}\,\pi_A \pi^A \right)} \right)^{1/2} + N^i\, \partial_i\phi^A \pi_A \;.
\end{equation}
This Hamiltonian has the form $\mathcal H = N\,\mathcal H_0 + N^i \,\mathcal H_i$, linear in the ADM lapse and shift, as it should be in any minimally coupled theory where the scalar fields enter the action only through different combinations of $g^{\mu\nu}\partial_\mu\phi^A\partial_\nu\phi^B$. When considered together with gravity, $\mathcal H_0$ and $\mathcal H_i$ simply contribute to the Hamiltonian constraint and to the generators of spatial diffeomorphisms respectively. Doing so does not change the dynamics in the scalar field sector, and merely reflects the reparametrization invariance of the action. We therefore feel free to consider the scalar fields separately from gravity. At last, we note also that the scalar fields Hamiltonian $\mathcal H_0$ can be written as a trace of the square root matrix $\mathcal H_0 = - \Tr \sqrt{S^A_C \left( \delta^C_B +\gamma^{-1} \, \pi^C \,\pi_B  \right)}$, very similar to the structure of the Lagrangian $\mathcal L_\phi=\sqrt{-g}\Tr\sqrt{\mathcal I}$.

In the case when $\beta_1 = 0$ and $\beta_2 = 1$ the velocities can be expressed as
\begin{equation}
D\phi^A = - \frac{\sqrt{\det\mathcal I}}{\sqrt{\gamma}\,\det S}\, S^{AB} \pi_B\;.
\end{equation}
Using the relation $\frac{\det S}{\det\mathcal I} = 1 + \frac{S^{AB} \pi_A\pi_B}{\gamma\,\det S}$ one can obtain the scalar fields Hamiltonian as
\begin{equation}
\mathcal H_0 = - \sqrt\gamma\, \sqrt{\det S + \gamma^{-1}\,\pi_A\pi_B S^{AB}} \equiv -\sqrt\gamma\, \det\sqrt{S^A_C \left( \delta^C_B +\frac{S^C_D \pi^D \,\pi_B}{\gamma\,\det S}  \right)}\;.
\end{equation}
The form of this Hamiltonian is also similar to the form of the original Lagrangian, $\mathcal L_\phi=~\sqrt{-g}\det \sqrt{\mathcal I}$, and can be written as a determinant of some square root matrix.

In order to look at the dynamics we focus on the former case with $\beta_2=0$. The equations of motion for the scalar fields in Hamiltonian form read
\begin{align}
D{\phi}^A &= \frac1{\mathcal H_0} \left( \frac{\det S}{\sqrt{\det S \, \left(1 + \gamma^{-1}\,\pi_A \pi^A \right)}} \, \pi^A + S^{AB}\pi_B \right)\;,\\
\dot{\pi}_A - \partial_i \left( N^i\, \pi_A \right)&= \partial_i \left(\frac{\gamma\,N}{\mathcal H_0} \left[ \eta_{AB} + \gamma^{-1}\,\pi_A\,\pi_B + \sqrt{\det S \, \left(1 + \gamma^{-1}\,\pi_A \pi^A \right)}\; S^{-1}_{AB}\right] \partial^i \phi^B \right) \label{eq:pieom},
\end{align}
where the inverse of the spatial derivative matrix is $S^{-1}_{AB}=(S\eta_{AB}-S_{AB})/\det S$. These equations of motion describe the evolution of the scalar fields $\phi^A$ and their conjugated momenta $\pi_A$ for any initial conditions with $\det S \neq 0$. In general one expects that all the regular solutions, i.e. the solutions specified with the initial conditions with $\det S \neq 0$, are tangential to the surface $\det S = 0$ at some point of time. In other words, the singular solutions, for which $\det S = 0$ at any time, are the envelopes of the families of regular solutions. Choosing the conditions in vicinity of the singular surface and following the infinitesimal evolution in time, one can study the phase portrait of the system near the singular surface and the way regular solutions are connected to the singular ones. We also note that the Hamiltonian~\eqref{eq:ham} cannot be used to study the singular solutions themselves since it relies on the assumption $\det S \neq 0$. The solutions with $\det S = 0$ shall therefore be studied separately.

\subsection{Time evolution in the vicinity of the singular surface}

\begin{figure}
\center \includegraphics{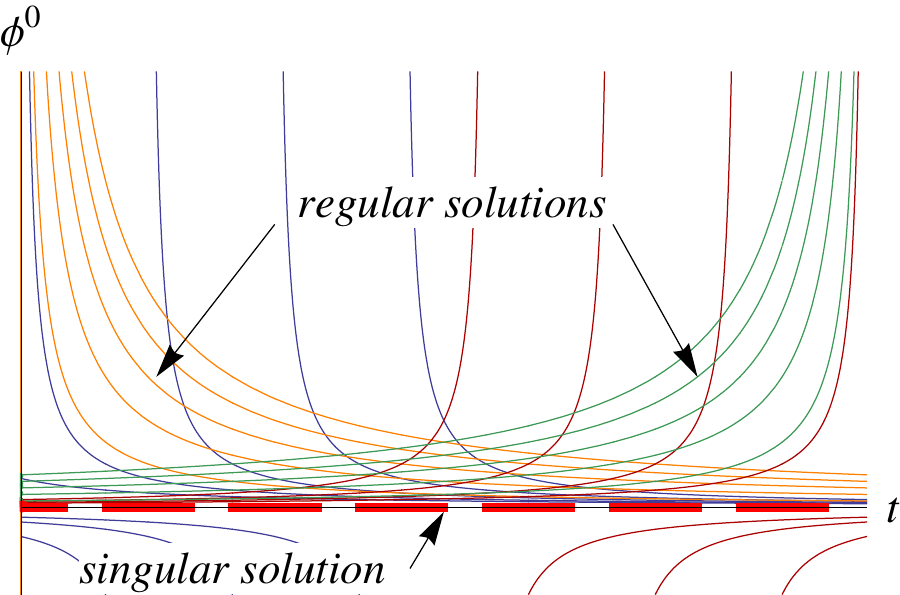}
\caption{Some members of the family of regular solutions~\eqref{eq:solution} (solid), which approach given singular solution $\phi^0 = 0$ (dashed) in the limit $t \to \pm\infty$.\label{fig:sol}}
\end{figure}

In order to illustrate the behaviour of regular solutions in the vicinity of the $\det S = 0$ surface let us choose some initial conditions that are infinitesimally close to the known trivial solution $\phi^0 = t$, $\phi^1 = x^1$, but have non-vanishing $\det S$ and its time derivative. Simplest way to write such initial conditions is to provide a small $x^2$ (or $x^3$) dependence to the $\phi^0$:
\begin{equation}\label{eq:init}
\phi^0(t_0) = t_0 + \eps_0 \, x^2\,,\quad \dot\phi^0(t_0) = 1 + \dot\eps_0\, x^2\,, \quad \phi^1(t_0) = x^1\,,\quad \dot\phi^1(t_0) = 0\,,
\end{equation}
where $\eps_0$, and $\dot\eps_0$ are arbitrary constants, vanishing in the case of the critical solution $\phi^A = x^A$. Since $\det S = - \eps_0^2$ and $\frac{d}{dt} \det S = - 2\dot\eps_0\eps_0$, these constants characterize the displacement from the singular surface and its time derivative at the initial moment. Using the Hamiltonian equation~\eqref{eq:pieom} for the momenta one can follow the infinitesimal evolution of fields $\phi^A$ in time. Moreover it happens to be possible to find an exact solution in the case of initial conditions~\eqref{eq:init}. It can be obtained by promoting the $x^2$ dependence of the initial conditions to be valid at all times. By plugging the ansatz $\phi^0(t) = \xi(t) + \eps(t) \, x^2$ into the equations of motion \eqref{eq:pieom} one obtains two equations for the functions $\xi(t)$ and $\eps(t)$
\begin{equation}
\ddot\xi(t) = 2\, \dot\xi(t) \, \frac{\dot\eps(t)}{\eps(t)} \,, \qquad \ddot\eps(t) =  2\, \frac{\dot\eps(t)^2}{\eps(t)}\,. 
\end{equation}
The general solution for the fields $\phi^A(t)$ is given by
\begin{equation}\label{eq:solution}
\phi^0(t) = \frac{\dot\xi_0\, (t-t_0) + \eps_0\,x^2}{1- \frac{\dot\eps_0}{\eps_0}(t - t_0)} + \xi_0\;, \qquad \phi^1 (t)= x^1\;.
\end{equation}
It happens that this family of solutions never approaches the singular trivial solution $\phi^0 = t$ independently of how close are the initial conditions to it, i.e. how small is $\eps_0$. Instead, the solutions~\eqref{eq:solution} are asymptotically approaching a different set of singular solutions $\phi^0 = \dot\xi_0\,\frac{\eps_0}{\dot\eps_0}+ \xi_0 = const$ in the limit $t \to \pm\infty$. Therefore, for any given constant there is a three parameter subfamily of regular solutions that approach it in the $t \to \pm\infty$ limit. Figure~\ref{fig:sol} illustrates this behaviour for the singular solution $\phi^0 = 0$. For simplicity we have suppressed the $x^2$ dependence of the regular solutions, and each line on the figure~\ref{fig:sol} corresponds to the one parameter family of solutions, which are different from each other by the constant rescaling of $\eps(t)$. From the solution~\eqref{eq:solution} one can see that $\det S \propto \frac1{t^2}$, and therefore the $\det S = 0$ surface cannot be reached along the discussed trajectory at any finite time. Note that apart from the solutions that start at the finite distance from the $\det S =0$ surface and approach it in the future there exist solutions that start as a small perturbations of the singular solution and leave the $\det S = 0$ surface. We would also like to remark that all the solutions with a given $x^2$ dependence have a singularity at the finite time $t = t_0 + \frac{\eps_0}{\dot\eps_0}$. Hopefully there are other solutions in this theory that are free of singularities.


\section{The singular surface with $\det S=0$}\label{sec:4}
In this section we discuss the most general scalar field configurations which satisfy the condition $\det S=0$ and show that the dynamics of the scalar fields in this subspace are equivalent to the dynamics of scalar fields in the case of $1+1$ space-time dimensions. To see this we first discuss the $1+1$ dimensional case separately.

\subsection{1+1 dimensions: massive gravity}
In $1+1$ dimensions the scalar field Lagrangian \eqref{act3} reduces to
\begin{equation}\label{act4}
\mathcal L_\phi=\beta_1\sqrt{-g}\left(\mathrm{Tr}\,\mathcal I+2\sqrt{\det \mathcal I}\right)^{1/2},
\end{equation}
since the $\beta_2$ term is a total derivative term. The above Lagrangian coincides with the dRGT mass term and was previously analyzed in \cite{deRham:2011rn,Kluson:2011aq}. Here we shall follow a different approach of Hamiltonian analysis which enables us to find the gauge symmetry of the scalar field action. We show that the scalar field action of non-linear massive gravity propagates no degrees of freedom in $1+1$ dimensions, in agreement with \cite{deRham:2011rn, Kluson:2011aq}.

We first note that in this case the determinant of the matrix $\mathcal I^A_B$ is a full square
\begin{equation}
\det \mathcal I^A_B=\frac{\det f}{\det g}\left( \frac{1}{2}\bar\varepsilon_{AB}\bar\varepsilon^{\mu\nu}\partial_\mu\phi^A\partial_\nu\phi^B\right)^2,
\end{equation}
where the bared Levi-Civita tensors $\bar\varepsilon^{\mu\nu},\,\bar\varepsilon_{AB}$ denote the flat space antisymmetric tensors defined as $\bar\varepsilon^{01}=-\bar\varepsilon^{10}=1$, etc. in every coordinate frame. In this section for simplicity we will consider the flat Minkowski  scalar field metric $f_{AB}=\eta_{AB}$ for which the factor $\det f =-1$. The scalar field action then becomes (up to a constant factor)
\begin{equation}
\mathcal L_\phi=2\frac{N}{\sqrt{\gamma^{11}}}\left[D\psi_++\sqrt{\gamma^{11}}\psi_+'\right]^{1/2}\left[D\psi_--\sqrt{\gamma^{11}}\psi_-'\right]^{1/2}\;,
\end{equation}
where $\psi_\pm\equiv\phi^0\pm\phi^1$, $D\equiv (\partial_0-N^1\partial_1)/N$, and $\psi_\pm'\equiv\partial_1\psi_\pm$. In the following we perform the full Hamiltonian analysis of this system according to the constraint analysis proposed by Dirac and extended by Henneaux et al. \cite{dirac, Henneaux:1990au}.

\subsection{Minkowski background}
We start with the case of a flat Minkowski background metric $g_{\mu\nu}=\eta_{\mu\nu}$ since the generalization to an arbitrary background is straightforward, as we shall see below.

\subsubsection{Constraint algebra}
In flat space the Lagrangian takes the simple form
\begin{equation}\label{actflat}
\mathcal L_\phi=2\sqrt{\dot\psi_++\psi_+'}\sqrt{\dot\psi_--\psi_-'}\;,
\end{equation}
and the conjugated momenta to the fields $\psi_\pm$ are 
\begin{equation}\label{pi}
 \pi_+=\frac{\sqrt{\dot\psi_--\psi_-'}}{\sqrt{\dot\psi_++\psi_+'}}\;,\qquad\pi_-=\frac{\sqrt{\dot\psi_++\psi_+'}}{\sqrt{\dot\psi_--\psi_-'}}\;.
 \end{equation}
It is obvious that the momenta are not independent. Instead, they satisfy the primary constraint
 \begin{equation}
 \mathcal C_0\equiv\pi_+-\frac{1}{\pi_-}=0\;.
 \end{equation}
 The total Hamiltonian density of the theory is obtained by adding the primary constraint to the Hamiltonian as
\begin{align}\label{tham}
\mathcal H_T&=\pi_+\dot\psi_++\pi_-\dot\psi_--\mathcal L_\phi+u_0\mathcal C_0\nonumber\\
&=\pi_-\psi_-'-\frac{1}{\pi_-}\psi_+'+u_0\mathcal C_0\;,
\end{align}
where the Lagrange multiplier $u_0=u_0(t,x)$ is an arbitrary function of the space-time coordinates. 
 For the analysis of the dynamics of the system we define the equal-time Poisson bracket as 
 \begin{equation}\label{pb}
 \{f(x),g(x')\}=\int dz\left(\frac{\delta f(x)}{\delta \psi^i(z)}\frac{\delta g(x')}{\delta\pi_i(z)}-\frac{\delta g(x')}{\delta \psi^i(z)}\frac{\delta f(x)}{\delta\pi_i(z)}\right).
 \end{equation}
 The time evolution of a functional $f(t,x)=f(t,x,\,\psi^i(t,x),\,\pi_i(t,x))$ is then given by
 \begin{equation}
 \frac{d}{dt}f(t,x)=\frac{\partial f(t,x)}{\partial t}+\int{dx'}\{f(t,x),\mathcal H_T(t,x')\}\;.
 \end{equation}
 For the consistency of the Hamiltonian equations of motion with the Lagrangian equations of motion one has to impose an additional constraint to the system, namely that the primary constraint is preserved in time. This in general leads either to secondary (and tertiary, $\dots$) constraints or determines the arbitrary function $u_0(t,x)$ \cite{dirac}. In our case we obtain a secondary constraint
 \begin{equation}
 \frac{d}{dt}\mathcal C_0(t,x)=-2\left(\frac{1}{\pi_-}\right)'\equiv 2\mathcal C_1(t,x)\;.
 \end{equation}
It is straightforward to check that the time evolution of $\mathcal C_1$ does not imply any new constraints since
\begin{equation}
\frac{d}{dt}\mathcal C_1(t,x)=\int{dx'}\{\mathcal C_1(t,x),\mathcal H_T(t,x')\}=-\left(\frac{1}{\pi_-}\right)''=\mathcal C_1'
\end{equation}
is a spatial derivative of the secondary constraint itself. Since both constraints mutually commute, i.e. $\{\mathcal C_0(x),\mathcal C_0(x')\}= \{\mathcal C_0(x),\mathcal C_1(x')\}=\{\mathcal C_1(x),\mathcal C_1(x')\}=0$, and since there are no further constraints, we conclude that the constraint algebra is closed and our system has two first class constraints. $\mathcal C_0$ is a primary first-class constraint and $\mathcal C_1$ is a secondary first-class constraint.

\subsubsection{Gauge symmetry}
The existence of first-class constraints indicates that there is a gauge symmetry in our theory. The purpose of this section is therefore to identify the gauge symmetries of the original Lagrangian \eqref{actflat} and find the number of degrees of freedom described by it.

Since the total Hamiltonian \eqref{tham} contains an arbitrary function of space-time coordinates $u_0$, a given set of initial conditions for the canonical variables $\psi^i,\,\pi_i$ after some time interval will evolve to different values of the canonical variables for different choices of $u_0$. Any two such set of values describe the same physical state related by a gauge transformation. In order to find the generators of the transformation one considers the evolution of a given set of initial data over a \emph{finite} time interval. This is reached by \emph{multiple} Poisson brackets of the canonical variables and total Hamiltonian, each of them transforming the system infinitesimally. Hence after a finite time interval two different sets of canonical variables obtained from the same initial data will differ by a gauge transformation generated by \emph{all} first-class constraints. It is therefore why all the first-class constraints should be put on the same footing and the Hamiltonian should be extended by adding to it also the secondary (and tertiary, $\dots$) first-class constraints \cite{dirac}. This makes the full symmetry of the theory manifest. In our case the extended Hamiltonian looks like
\begin{equation}\label{eham}
\mathcal H_E=\pi_-\psi_-'-\frac{1}{\pi_-}\psi_+'+u_0\mathcal C_0+u_1\mathcal C_1\;,
\end{equation}
where we have introduced another arbitrary function $u_1(t,x)$. Under the transformations generated by the two constraints the canonical variables $q=\left\{\psi^i,\,\pi_i\right\}$ transform according to the law
\begin{align}\label{trafo1}
&q\mapsto q+\delta q\;,\qquad\delta q(x)=\left\{q(x),\int dx'\left[\varepsilon_0(x')C_0(x')+\varepsilon_1(x')C_1(x')\right]\right\}.
\end{align}
This for the transformations of the canonical fields gives
\begin{align}\label{trafo2}
\delta\psi_+=\varepsilon_0\;,\qquad\delta\psi_-=\frac{1}{\pi_-^2}(\varepsilon_0-\varepsilon_1')\;,
\end{align}
while the conjugated momenta stay unchanged. The corresponding \emph{extended} first order action
\begin{equation}\label{acte}
S_E=\int d^2x\left[\pi_+\dot\psi_++\pi_-\dot\psi_--\mathcal H_E\right]
\end{equation}
is invariant under the above gauge transformations if also the Lagrange multipliers $u_0,\,u_1$ transform. Their transformation laws are not of any need in the present work, therefore we shall not give their explicit form and instead refer the reader to \cite{Henneaux:1990au}. Due to the fact that in \eqref{acte} we have introduced an additional arbitrary function $u_1$, the equations of motion which can be derived from \eqref{acte} do not coincide with the equations of motion following from the action $S_T=\int d^2x\left(\pi_i\dot\psi^i-\mathcal H_T\right)$ or equivalently from the original action \eqref{actflat}. Moreover, the original Lagrangian is not invariant under the gauge transformations \eqref{trafo2}. The reason for this is that the extended Hamiltonian formalism introduces an additional redundancy in the description. However, the time evolution of the gauge invariant fields can be equally well described by both the total Hamiltonian $\mathcal H_T$ and the extended Hamiltonian $\mathcal H_E$.

In order to obtain the symmetry of the original scalar field action, one can rewrite the transformations \eqref{trafo2} by expressing the conjugated momenta according to their definitions \eqref{pi} and demand that the action remains unchanged. This leads to the following relation between the gauge parameters
\begin{equation}
\varepsilon_0=\frac{1}{2}\left(\varepsilon_1'-\dot\varepsilon_1\right)\;.
\end{equation}
Hence the gauge symmetry of the Lagrangian is
\begin{align}
&\psi_-\mapsto\psi_--\frac{1}{2}(\varepsilon'+\dot\varepsilon)\frac{\dot\psi_--\psi_-'}{\dot\psi_++\psi_+'}\;,\\
&\psi_+\mapsto\psi_++\frac{1}{2}(\varepsilon'-\dot\varepsilon)\;.
\end{align}
Since the above symmetry transformation involves both, the gauge parameter $\varepsilon$ and its time derivative, then the number of degrees of freedom in the theory are reduced by two which coincides with the total number of first class constraints \cite{Henneaux:1990au}. It is so, because the gauge parameter and its time derivatives are independent functions in the sense of independent initial data which can be chosen arbitrarily at the initial moment of time.~\footnote{A familiar example where exactly the same approach of counting the degrees of freedom can be applied is electrodynamics. There the gauge transformation of the vector field $A_\mu\to A_\mu+\partial_\mu\lambda$ also involves both the gauge parameter $\lambda$ and its time derivative. The constraint analysis of the theory also shows that there is one primary and one secondary first-class constraint removing two out of four degrees of freedom. } Another way to see that there are no propagating degrees of freedom is by performing the gauge fixing in the extended action \eqref{acte}. Since there are two constraints on the momenta and two gauge symmetries \eqref{trafo2} on the canonical fields it is evident that the action is pure gauge and propagates no degrees of freedom. The same conclusion could have been drawn also from the analysis of the Lagrangian equations of motion.

\subsection{Arbitrary background}\label{sec:4.3}
The scalar field action in an arbitrary curved $1+1$ dimensional space-time can be written as
\begin{equation}\label{actcurv}
S_\phi=2\int d^2 x\,\sqrt{\gamma_{11}}\left[\dot\psi_++a_+\psi_+'\right]^{1/2}\left[\dot\psi_--a_-\psi_-'\right]^{1/2}\;,
\end{equation}
where we have introduced the notations $a_\pm=N\sqrt{\gamma^{11}}\mp N^1$. The conjugated momenta are defined as
\begin{equation}
 \pi_+=\sqrt{\gamma_{11}}\frac{\sqrt{\dot\psi_--a_-\psi_-'}}{\sqrt{\dot\psi_++a_+\psi_+'}}\;,\qquad\pi_-=\sqrt{\gamma_{11}}\frac{\sqrt{\dot\psi_++a_+\psi_+'}}{\sqrt{\dot\psi_--a_-\psi_-'}}\;,
 \end{equation}
and the Hamiltonian analysis of the system can be carried out in complete analogy to the case of Minkowski background. The extended Hamiltonian and the closed set of constraints can be found to be
\begin{align}
&\mathcal H_E= a_-\pi_-\psi_-'-a_+\frac{\gamma_{11}}{\pi_-}\psi_+'+u_0\mathcal C_0+u_1\mathcal C_1\;,\\
&\textrm{with }\quad \mathcal C_0=\pi_+-\frac{\gamma_{11}}{\pi_-}\;,\quad \mathcal C_1=-\left(\frac{1}{\pi_-}\right)'\;.
\end{align}
As before the constraints $\mathcal C_0$ and $\mathcal C_1$ are first class constraints and generate the gauge transformations of the canonical variables $\psi_\pm\mapsto\psi_\pm+\delta\psi_\pm$ with
\begin{align}\label{trafo3}
\delta\psi_+=\varepsilon_0\;,\qquad\delta\psi_-=\frac{1}{\pi_-^2}(\gamma_{11}\varepsilon_0-\varepsilon_1')\;.
\end{align} 
By inserting them in the Lagrangian \eqref{actcurv} one obtains the following condition on the gauge variables
\begin{equation}\label{trafo4}
\varepsilon_0'\gamma_{11}(a_++a_-)-\varepsilon_0\left(\partial_0-a_-\partial_1-2a_-'\right)\gamma_{11}+\left(\partial_0-a_-\partial_1-2a_-'\right)\varepsilon_1'=0\;,
\end{equation}
under which the Lagrangian remains invariant under the transformations \eqref{trafo3}. This condition can be rewritten in metric components by using the relations $\gamma_{11}=g^{00}\det g$, $\gamma_{11}(a_++a_-)=2\sqrt{-g}$, and
\begin{equation}
a_{\pm}=\frac{1}{g^{00}}\left(\varepsilon^{01}\pm g^{01}\right)\;,\qquad \partial_0\pm a_\pm\partial_1=\frac{1}{g^{00}}\left(g^{0\mu}\pm\varepsilon^{0\mu}\right)\partial_\mu\;,
\end{equation}
where the only non-zero components of the Levi-Civita tensor are $\varepsilon^{01}=-\varepsilon^{10}=-(\sqrt{-g})^{-1}$. Unfortunately, for generic background metric it is impossible to solve \eqref{trafo4} for $\varepsilon_0$ in local form. The gauge transformation is therefore in general non-local.

\subsection{3+1 dimensions}

For the two scalar fields in 3+1 dimensions the determinant of the matrix of spatial derivatives $S^{AB}\equiv\gamma^{ij}\partial_i\phi^A\partial_j\phi^B$ reads
\begin{align}
\det S^{AB}&=\left[\varepsilon^{ijk}\partial_j\phi^{0}\partial_k\phi^{1}\right]\gamma_{il}\left[{\varepsilon}^{lmn}\partial_m\phi^{0}\partial_n\phi^{1}\right]\;.
\end{align}
Hence the condition $\det S=0$ translates into requirement that the norm of the cross product of the spatial gradients of the scalar fields $\phi^0$ and $\phi^1$ vanishes. In other words it means that both gradients of the scalar fields have to lie along the same spatial direction and thus can be used to parametrize only one spatial direction. Therefore the most general scalar field configuration satisfying $\det S=0$ can be parametrized as 
\begin{align}\label{sc1}
\phi^0=\phi^0(t,\,f(t,x^i))\;,\\\label{sc2}
\phi^1=\phi^1(t,\,f(t,x^i))\;,
\end{align}
where $f(t,x^i)$ is an arbitrary function of space-time coordinates.

In order to see that this ansatz for the scalar fields makes the dynamics of the $3+1$ dimensional theory equivalent to the dynamics of the $1+1$ dimensional theory it is useful to introduce the short hand notations $N=\tilde N,\quad N^i\partial_if-\partial_0f=\tilde N^f,\quad \partial_if\partial_jf\gamma^{ij}=\tilde \gamma^{ff}$. In terms of these variables the $3+1$ dimensional field $\mathcal I^{AB}$ takes the form
\begin{align}
\mathcal I ^{AB}&\equiv g^{\mu\nu}\partial_\mu\phi^A\partial_\nu\phi^B=-\frac{1}{\tilde N^2}\left(\partial_t-\tilde N^f\partial_f\right)\phi^A\left(\partial_t-\tilde N^f\partial_f\right)\phi^B+\tilde\gamma^{ff}\partial_f\phi^A\partial_f\phi^B\nonumber\\
&=\mathcal I^{AB}_{(2)}\equiv \tilde g^{\tilde\mu\tilde\nu}\partial_{\tilde\mu}\phi^A\partial_{\tilde\nu}\phi^B\;,
\end{align}
where the tilded indices take the values $\tilde\mu=0,f$. We recognize the tilded variables $\tilde N,\,\tilde N^f,\,\tilde \gamma^{ff}$ as the ADM variables of an effective $1+1$ dimensional metric $\tilde g^{\tilde \mu\tilde\nu}$. Indeed, for the components of the effective metric 
\begin{align*}
\tilde g^{00} &=g^{tt}\;, \quad\tilde g^{0f}=g^{ti}\partial_if+g^{tt}\partial_tf\;, \\
\tilde g^{ff}&=g^{tt}\partial_tf\partial_tf+2g^{ti}\partial_tf\partial_if+g^{ij}\partial_if\partial_jf\;,
\end{align*}
 they satisfy
\begin{equation}\label{adm}
\tilde g^{00}=-\frac{1}{\tilde N^2}\;,\qquad \tilde g^{0f}=\frac{\tilde N^f}{\tilde N^2}\;,\qquad \tilde g^{ff}=\tilde\gamma^{ff}-\left(\frac{\tilde N^f}{\tilde N}\right)^2\;.
\end{equation}
As in the $1+1$ dimensional case, the determinant $\det \mathcal I$ can be rewritten as a full square
\begin{align}
\det\mathcal I\equiv\det\mathcal I^A_B=\frac{\det f}{\det\tilde g}\left[\frac{1}{2}\,\bar\varepsilon_{AB}\bar\varepsilon^{\tilde\mu\tilde\nu}\partial_{\tilde\mu}\phi^{A}\partial_{\tilde\nu}\phi^{B}\right]^2
\end{align}
with $\left(\det\tilde g\right)^{-1}=-\tilde \gamma^{ff}/\tilde N^2$, and $\bar\varepsilon^{\tilde\mu\tilde\nu},\,\bar\varepsilon_{AB}$ denoting the flat space antisymmetric tensors. Hence all the terms in the Lagrangian containing the scalar fields can be rewritten in terms of an effective two-dimensional metric $\tilde g^{\tilde\mu\tilde\nu}$. We would like to emphasize that this rewriting is merely cosmetic and has the meaning only as the simplification of notations in the scalar field action.

In order to simplify the analysis of the equations of motion of the scalars, we rewrite the integration measure of the Lagrangian density in another coordinate system $\{\tilde x^\mu\}$, where $\tilde x^0=t,\,\tilde x^1=f(t,x^i)$, and $\tilde x^2=\tilde x^2(x^\mu),\,\tilde x^3=\tilde x^3(x^\mu)$ some arbitrary non-degenerate coordinate transformations. In this case the metric components transform according to the usual transformation laws, and the components $\tilde g^{00},\,\tilde g^{01},\,\tilde g^{11}$ coincide with the components of the effective $1+1$ dimensional metric $\tilde g^{\tilde\mu\tilde\nu}, \,\tilde \mu=\{0,f\}$ given above. Hence the Lagrangian of the scalar fields can be rewritten in terms of the metric $\tilde g^{\mu\nu}$ as
\begin{equation}
S_\phi=2\int d\tilde x^2\,d\tilde x^3\int dt\,df\,\sqrt{-\tilde g}\frac{1}{\tilde N}\left[\dot\psi_++\tilde a_+\partial_f\psi_+\right]^{1/2}\left[\dot\psi_--\tilde a_-\partial_f\psi_-\right]^{1/2}
\end{equation}
where $\tilde a_\pm=\tilde N\sqrt{\tilde\gamma^{ff}}\mp \tilde N^f$. The variables $\tilde N,\,\tilde N^f,\,\tilde \gamma^{ff}$, used for notational simplicity only, can be expressed in terms of the metric $\tilde g^{\mu\nu}$ as in \eqref{adm}. By comparing this action with \eqref{actcurv} one sees that the only difference is the volume factor and the prefactor $\sqrt{-\tilde g}/\tilde N\neq\sqrt{\gamma_{11}}$, which depends on all four space-time coordinates. Under the assumption that the volume spanned by $\tilde x^2,\,\tilde x^3$ is finite, the Hamiltonian analysis of the scalar field dynamics coincides with that in section \ref{sec:4.3}.

We thus conclude that the ansatz for the scalar fields \eqref{sc1}, \eqref{sc2} such that the condition $\det S=0$ is satisfied leads to a theory which is equivalent to the $1+1$ dimensional case and thus propagates no degrees of freedom. In Hamiltonian language, on this subspace of the scalar field configurations the theory has two first class constraints.

\section{Conclusions}\label{sec:5}

Any diffeomorphism invariant formulation of massive gravity inevitably contains a number of scalar fields minimally coupled to the dynamical metric field and can be viewed as just some particular scalar field theory coupled to general relativity. Therefore we argue that the Hamiltonian structure and the counting of degrees of freedom can be done for gravity and scalar fields separately. In other words, the absence of the sixth degree of freedom in the dRGT non-linear massive gravity \cite{grt} can be seen as a feature of the scalar field action, and can be studied in the scalar field theory given by the dRGT mass term.

While the full dRGT scalar action contains the number of fields equal to the space-time dimension, in this paper we have focused on the reduced case with two scalar fields, which coincides with the full theory only in $1+1$ dimensions. We have calculated the determinant of the kinetic matrix $\partial^2\mathcal L_\phi/\partial\dot\phi^A\partial\dot\phi^B$ of the non-linear theory and have found that in $d>1$ dimensions it does not vanish for generic initial conditions. Thus in more than $1+1$ dimensions both of the fields are, in general, propagating. However there exists a subspace of the configuration space where the Hessian is vanishing. It corresponds to the case where the coordinate transformation represented by the scalar fields $\phi^A(x^i)$ is singular on any two-dimensional space-like surface, or, equivalently, when both of the fields depend only on one independent space-like direction. In this case the scalars effectively live on the $1+1$ dimensional space-time, and the theory is equivalent to the $1+1$ dimensional dRGT massive gravity, where there is only single spatial direction available. For the latter constrained theory the full Hamiltonian analysis reveals two first-class constraints which generate one gauge transformation that leaves the action invariant. Since the transformation involves two independent parameters, then after fixing the gauge the theory does not contain any degrees of freedom. This is in agreement with the previous findings in the $1+1$ dimensional dRGT massive gravity~\cite{deRham:2011rn,Kluson:2011aq}. For the theory in more than $1+1$ dimensions the effectively $1+1$ dimensional solutions with vanishing Hessian correspond to the so-called~\emph{singular solutions}. On such a singular solution at each moment in time there exist infinitely many other regular solutions of the theory which are tangential to the singular solution, i.e. with coinciding $\phi^A(x^i)$ and $\dot\phi^A(x^i)$. Therefore, there is no choice of initial conditions that uniquely specifies such a solution, and any perturbation in the initial conditions leads to the regular solution with two degrees of freedom and non-vanishing Hessian. We note that our findings do not allow us to draw conclusions about the behaviour of the dRGT-like theories with more than two scalar fields, but the proposed method can be extended to include arbitrary number of scalar fields.

\acknowledgments
We are grateful to Viatcheslav Mukhanov and Claudio Bunster for valuable discussions. It is also a pleasure to thank Alexander Vikman and Sergey Sibiryakov for their helpful comments on the manuscript. LA is supported by the DFG Cluster of Excellence EXC 153 ``Origin and Structure of the Universe''. AK is supported by Alexander von Humboldt Foundation.

\end{document}